\documentstyle[epsfig]{aipproc}

\begin{document}

\title{Large \( B- \)Fields and Noncommutative Solitons%
\thanks{This work is dedicated to my parents Juan Moreno and Martha Soto.
} }

\author{Jorge Moreno%
\thanks{E-mail address: jmoreno@fis.cinvestav.mx
}}

\address{Departamento de F\'{\i}sica\\
Centro de Investigaci{\'o}n y de Estudios Avanzados del I.P.N.\\
Apdo. Postal 14-740, 07000, M{\'e}xico D.F. M{\'e}xico}

\maketitle

\begin{abstract}
The purpose of this talk is to review a few issues concerning noncommutativity
arising from String Theory. In particular, it is shown how in Type
IIB Theory, the annihilation of a \( D3-\overline{D3} \) brane pair
yields a \( D1- \)string. This object, in the presence of a large
\( B- \)field and fermions, happens to be a complex noncommutative
soliton endowed with superconductivity.
\end{abstract}

\section*{INTRODUCTION}

Type IIB Superstring Theory allows two interesting ingredients: stable-BPS
RR-charged \( Dp \)-branes (\( p \) odd) and a massless antisymmetric
\( B_{\mu \nu } \) field. Although a \( D3-D3 \) brane system is
stable; a \( D3-\overline{D3} \) case is not, due to the presence
of a tachyon in its spectrum\cite{key-1}. The job of the \( B \)-field
in the low energy limit is to introduce noncommutativity\cite{key-4}.
In the \( D3-\overline{D3} \) brane configuration, we may turn on
a large \( B \)-field along two worldvolume spatial coordinates.
The effect of this is that the complex tachyon allows a GMS soliton\cite{Seiberg}.
This object appears to be superconducting in the presence of fermions
arising from the open string sector\cite{Moreno}.

\section*{\protect\( B\protect \)-FIELDS AND NONCOMMUTATIVITY}

Consider an open string attached to a \( D \)-brane. The OPE has
the form\begin{equation}
\label{OPE}
e^{ik_{1}\cdot X}e^{ik_{2}\cdot X}\sim \left( \tau -\tau '\right) ^{2\alpha 'g^{\mu \nu }k_{1\mu }\cdot k_{2\nu }}\times e^{i\left( k_{1}+k_{2}\right) \cdot X}+\cdots .
\end{equation}
However, the introduction of a large \( B \)-field alters this to
\begin{equation}
\label{OPEB}
e^{ik_{1}\cdot X}e^{ik_{2}\cdot X}\sim \left( \tau -\tau '\right) ^{2\alpha 'G^{\mu \nu }k_{1\mu }\cdot k_{2\nu }}\times \left[ e^{-\frac{i}{2}\Theta ^{\mu \nu }k_{1\mu }\cdot k_{2\nu }}\right] \times e^{i\left( k_{1}+k_{2}\right) \cdot X}+\cdots ,
\end{equation}
 where \( G^{\mu \nu }=\left( \frac{1}{g+2\pi \alpha 'B}g\frac{1}{g-2\pi \alpha 'B}\right) ^{\mu \nu } \)
is the effective metric seen by the open string modes, and \( \Theta ^{\mu \nu }=-\left( 2\pi \alpha '\right) ^{2}\left( \frac{1}{g+2\pi \alpha 'B}B\frac{1}{g-2\pi \alpha 'B}\right) ^{\mu \nu } \)
is the noncommutativity parameter matrix\cite{Seiberg}. Likewise,
the new term in brackets is known -in configuration space- as the
Moyal \( * \) product. Thus, in the low energy effective theory,
fields get \( * \)-multiplied.

\section*{NONCOMMUTATIVE SOLITONS}

The idea of GMS solitons was cleverly used in \cite{Harvey} to construct
real solutions to the tachyon in bosonic \( D \)-branes. In this
work, complex tachyons in RR \( D \)-brane pairs are considered instead.
For a \( D3-\overline{D3} \) brane pair in the presence of a large
\( B \)-field along the \( x-y \) plane, the action of the tachyon
is\begin{equation}
\label{tachyon}
S^{ \left( \Sigma _{3+1} \right) }=\int _{\Sigma _{3+1}}dxdydzdt\left[ \overline{D^{\mu }T}*D_{\mu }T-V_{*}\left( T,\overline{T}\right) \right] ,
\end{equation}
 where \( \left[ x,y\right] =i\theta , \) and \( z \) and \( t \)
are commutative coordinates. Also, \( D_{\mu }T=\partial _{\mu }T-iA_{\mu }*T+i\widetilde{A_{\mu }}*T, \)
where \( A_{\mu } \) and \( \widetilde{A_{\mu }} \) are the respective
gauge fields in each of the \( D \)-brane's Chan-Paton \( U\left( 1\right)  \)
symmetry groups.

We make three assumptions:

\begin{itemize}
\item The potential is a polynomial: \begin{equation}
\label{potential}
V_{*}\left( T,\overline{T}\right) =\sum ^{n}_{k=1}a_{k}\left( \overline{T}*T\right) ^{k}.
\end{equation}

\item The gauge field \( R_{\mu }\equiv A_{\mu }-\widetilde{A_{\mu }} \)
has the form:\begin{equation}
\label{R}
R_{\mu }=R_{\mu }\left( z,t\right) .
\end{equation}

\item We'll focus on solutions of the form:\begin{equation}
\label{T}
T=T\left( x,y\right) ,\; \overline{T}=\overline{T}\left( x,y\right) .
\end{equation}

\end{itemize}
Eventually, it is shown that the solution is\begin{equation}
\label{solution}
T=t_{*}T_{0},\; \overline{T}=\overline{t_{*}}\overline{T_{0}},
\end{equation}
 where \( t_{*}\) and \( \overline{t_{*}} \) solve the equations of motion
in the commutative case, while \( T_{0},\overline{T_{0}} \) satisfy
\begin{equation}
\label{K}
\left( \overline{T_{0}}*T_{0}\right) ^{k}=\overline{T_{0}}*T_{0}.
\end{equation}

Since \( \left[ x,y\right] =i\theta  \) is analogous to \( \left[ \hat{q},\hat{p}\right] =i \)
in quantum mechanics, we identify \( T_{0}\leftrightarrow i\left| 0\right\rangle \left\langle 0\right|  \)
and \( \overline{T_{0}}\leftrightarrow -i\left| 0\right\rangle \left\langle 0\right|  \)
in the Simple Harmonic Oscillator basis.

Applying the Weyl-Wigner-Moyal correspondence yields the following
lowest energy solitonic solution: \begin{equation}
\label{Jorge}
T\left( x,y\right) =2it_{*}e^{-r^{2}},\; \overline{T}\left( x,y\right) =-2i\overline{t_{*}}e^{-r^{2}},
\end{equation}
 where \( r^{2}=x^{2}+y^{2} \)\cite{Moreno}.

\section*{THE NONCOMMUTATIVE SUPERCONDUCTING STRING}

Given a \( \left( 3+1\right)  \) Dirac spinor \( \Psi  \), with
two-component entries \( \psi _{R} \) and \( \psi _{L} \) obeying
\( \vec{\sigma }\cdot \widehat{p}\psi _{R}=\psi _{R} \) and \( \vec{\sigma }\cdot \widehat{p}\psi _{L}=-\psi _{L} \),
the action for the spinor coupled to the complex soliton has the form:\begin{equation}
\label{spinor}
S^{\left( \Sigma _{3+1}\right) }=\int _{\Sigma _{3+1}}dtdzdxdy\left[ f\left( \overline{T}\right) *\overline{\Psi }*g\left( T\right) D\! \! \! \! /\, \Psi \right] ,
\end{equation}
 where \( f \) and \( g \) are polynomials, and \( D\! \! \! \! /\, \Psi =\gamma ^{\mu }\left( \partial _{\mu }\Psi -iR_{\mu }*\Psi \right) . \)

In terms of operators, we may express our spinors as\cite{Pilo}:\begin{equation}
\label{operator}
\widehat{\psi }_{L,R}\left( x^{\mu }\right) =\sum _{m,n\geq 0}\psi ^{L,R}_{mn}\left( z,t\right) \left| m\right\rangle \left\langle n\right| .
\end{equation}

In order to find the effective theory along the \emph{string} (the
\( z,t \) coordinates,) we make two assumptions:

\begin{itemize}
\item \( \theta \longrightarrow \infty  \), which means that the noncommutative
kinetic part is negligible.
\item \( \psi \longrightarrow \psi ^{L} \) (as in Witten's superconducting
string \cite{Witten}).
\end{itemize}
Therefore, the action for the \emph{Noncommutative \( D \)-string} is\begin{equation}
\label{string}
S^{\left( \Sigma _{1+1}\right) }=-2\pi i\theta f\left( \overline{t_{*}}\right) g\left( t_{*}\right) \int _{\Sigma _{1+1}}\left[ i\overline{\psi }^{L}\sigma ^{a}D_{a}\psi ^{L}-\overline{\psi }^{L}m\psi ^{L}\right] ,
\end{equation}
 where \( \psi ^{L} \) denotes \( \psi ^{L}_{00} \) in (\ref{operator}),
\( m \) is a {}``mass'' matrix (\( m\left( z,t\right) =\sigma ^{\alpha }R_{\alpha } \))%
\footnote{\( a=z,t \) and \( \alpha =x,y \).
}.

\section*{SUPERCONDUCTIVITY }

In the massless case (after rescaling the action and getting rid of
the unnecessary \( L \) subscript)\begin{equation}
\label{effective}
S^{ \left( \Sigma _{3+1} \right) }=\int _{\Sigma _{3+1}} dzdt\left( i\overline{\psi }\sigma ^{a}D_{a}\psi \right) .
\end{equation}

According to the bosonization technique \cite{Witten}, in two dimensions
we can equivalently describe the theory by either bosons or fermions.

This is done by introducing a scalar field \( \zeta \left( z,t\right)  \)
such that \begin{equation}
\label{bosonization}
\overline{\psi }\sigma ^{a}\psi =\frac{1}{\sqrt{\pi }}\varepsilon ^{ab}\partial _{b}\zeta .
\end{equation}
 It can be shown that the kinetic term corresponds to \begin{equation}
\label{kinetic}
i\overline{\psi }\sigma ^{a}D_{a}\psi =\frac{1}{2}\left( \partial _{a}\zeta \right) \left( \partial ^{a}\zeta \right) -\frac{1}{\sqrt{\pi }}R_{a}\varepsilon ^{ab}\partial _{b}\zeta ,
\end{equation}
 which is associated to a conserved current \begin{equation}
\label{current}
J^{a}=\partial ^{a}\zeta +\frac{1}{\sqrt{\pi }}\varepsilon ^{ab}R_{b}.
\end{equation}
 However, this current may be expressed in terms of another scalar:\begin{equation}
\label{phi}
J^{a}=\varepsilon ^{ab}\partial _{b}\varphi .
\end{equation}
 Thus, \( \partial _{b}\varphi =-\varepsilon _{ba}J^{a} \) and \begin{equation}
\label{zfi}
\partial ^{b}\partial _{b}\varphi =-\partial ^{b}\varepsilon _{ba}\left( \partial ^{a}\zeta +\frac{1}{\sqrt{\pi }}\varepsilon ^{ab}R_{b}\right) =-\frac{1}{\sqrt{\pi }}\partial ^{b}R_{b}.
\end{equation}
 This means that from \( \partial _{a}\varphi =-\varepsilon _{ac}J^{c}, \)
we get that \( -\varepsilon _{ac}\partial ^{a}J^{c}=-\frac{1}{\sqrt{\pi }}\partial ^{a}R_{a}. \)
In other words, for \( J^{3} \) (the current along the string):\begin{equation}
\label{J3}
\frac{dJ^{3}\left( z,t\right) }{dt}=\frac{1}{\sqrt{\pi }}\frac{dR_{0}\left( z,t\right) }{dz},
\end{equation}
 which has the following solution:\begin{equation}
\label{superconducting}
J^{3}\left( z,t\right) =\frac{1}{\sqrt{\pi }}\left[ R_{0}\left( z,\tau \right) -R_{0}\left( z,\tau _{i}\right) \right] .
\end{equation}

This means that the current is nondecaying as long as the gauge fields
has different values when it's turn on (\( t=\tau _{i} \)) and turn
off (\( t=\tau  \)).

\section*{ACKNOWLEDGMENTS}

It is a pleasure to thank my advisor Dr. H. Garc\'{\i}a-Compe{\'a}n for useful
discussions and enjoyable collaboration. I'm very grateful to J.M.L.
Bauche A. for helping me write this manuscript. I also thank the organizers
of the \emph{VIII Mexican Workshop on Particles and Fields} for their hospitality.
This work was supported in part by the CONACyT Fellowship No. 33951E.


\begin{thebibliography}{1}
\bibitem{key-1}J. H. Schwarz, {}``TASI Lectures on Non-BPS D-Brane Systems'', hep-th/9908144.
\bibitem{key-4}R. Gopakumar, S. Minwalla, and A. Strominger, {}``Noncommutative
Solitons,'' JHEP \textbf{0005} (2000) 020, hep-th/0003160.
\bibitem{Seiberg}N. Seiberg and E. Witten, {}``String Theory and Noncommutative Geometry,''
JHEP \textbf{09} (1999) 032, hep-th/9908142.
\bibitem{Moreno}H. Garc\'{\i}a-Compe{\'a}n and J. Moreno, {}``Remarks on Noncommutative Solitons,''
hep-th/0110119.
\bibitem{Harvey}J. Harvey, P. Kraus, F. Larsen, and E.J. Martinec, {}``D-branes and
Strings as Noncommutative Solitons,'' JHEP \textbf{0007} (2000) 042,
hep-th/0005031.
\bibitem{Pilo}L. Pilo and A. Riotto, {}``The Non-commutative Brane World,'' JHEP
\textbf{03} (2001) 015, hep-ph/0012174
\bibitem{Witten}E. Witten, {}``Superconducting Strings,'' Nucl. Phys. B \textbf{249}
(1985) 557.\end{thebibliography}
\end{document}